\documentclass[twocolumn,showpacs,amsfonts,aps,prc,floatfix,%
               nofootinbib]{revtex4}

\usepackage{amsmath}
\usepackage{bm}
\usepackage{graphicx}

\newcommand{\Tdec}{T_\mathrm{dec}}
\newcommand{\Tc}{T_\mathrm{c}}
\newcommand{\Tchem}{T_\mathrm{chem}}

\newcommand{\VISH}{{\tt VISH2{+}1}}
\newcommand{\VC}{{\tt VISHNU}}

\newcommand{\ecc}{\varepsilon}
\newcommand{\dNdy}{dN_\mathrm{ch}/dy}

\newcommand{\be}[1]{\begin{equation}\label{#1}}
\newcommand{\ee}{\end{equation}}

\newcommand{\eq}{{\,=\,}}

\def\La{\langle}
\def\Ra{\rangle}
\def\lda{\langle\!\langle}
\def\rda{\rangle\!\rangle}

\usepackage{color}

\begin{document}

\title{Hydrodynamic flow in heavy-ion collisions with large hadronic viscosity}
\date{\today}

\author{Chun Shen}
\email[Correspond to\ ]{shen@mps.ohio-state.edu}
\affiliation{Department of Physics, The Ohio State University, 
  Columbus, OH 43210-1117, USA}
\author{Ulrich Heinz}
\email[Email:\ ]{heinz@mps.ohio-state.edu}
\affiliation{Department of Physics, The Ohio State University, 
  Columbus, OH 43210-1117, USA}

\begin{abstract}
Using the (2+1)-dimensional viscous hydrodynamic code {\tt VISH2+1} 
with a temperature dependent specific shear viscosity $(\eta/s)(T)$,
we present a detailed study of the influence of a large hadronic shear 
viscosity and its corresponding relaxation time $\tau_\pi$ on the 
transverse momentum spectra and elliptic flow of hadrons produced in 
$200\,A$\,GeV Au+Au collisions. Although theory, in principle, 
predicts a well-defined relation $\tau_{\pi}T\eq\kappa(T)\times
\left(\eta/s\right)(T)$, the precise form of $\kappa(T)$ for the matter 
created in relativistic heavy-ion collisions is not known. For the popular 
choice $\kappa\eq3$ the hadron spectra are found to be insensitive to a 
significant rise of $\eta/s$ in the hadronic stage, whereas their 
differential elliptic flow $v_2(p_T)$ is strongly suppressed by large 
hadronic viscosity. The large viscous effects on $v_2$ are 
strongly reduced if (as theoretically expected) $\kappa(T)$
is allowed to grow with decreasing temperature in the hadronic stage.
This implies that, until reliable calculations of $\kappa(T)$ become
available, an extraction of the hadronic shear viscosity from a comparison 
between {\tt VISH2+1} and a microscopic hadron cascade or experimental data 
requires a simultaneous fit of $(\eta/s)(T)$ and $\kappa(T)$.
\end{abstract}


\maketitle

\section{Introduction}
\label{sec1}

A fluid state of matter, quark-gluon plasma (QGP), is created in 
ultra-relativistic heavy-ion collision experiments at the Relativistic
Heavy Ion Collider (RHIC) \cite{Arsene:2004fa,Back:2004je,Adams:2005dq,%
Adcox:2004mh}. Theoretical analysis of these experiments established 
that QGP behaves like an almost perfect liquid with very small viscosity
\cite{Heinz:2001xi,Kolb:2003dz,Gyulassy:2004zy}. Much effort has been 
focused on determining the QGP transport parameters, in particular 
its specific shear viscosity $(\eta/s)_\mathrm{QGP}$, i.e. the ratio 
between its shear viscosity $\eta$ and entropy density $s$ (see 
\cite{Romatschke:2009im,Heinz:2009xj} for recent reviews). 

In a recent article \cite{Song:2010mg} the newly developed hybrid code 
{\tt VISHNU} \cite{Song:2010aq} has been used to extract 
$(\eta/s)_\mathrm{QGP}$ from the observed collision centrality dependence 
of the integrated charged hadron elliptic flow $v_2$. This code couples 
the macroscopic evolution of the QGP by the (2+1)-dimensional viscous 
hydrodynamic code {\tt VISH2{+}1} to the Boltzmann cascade {\tt UrQMD} 
which describes the final hadronic rescattering and freeze-out stage 
microscopically. The microscopic simulation of the late hadronic stage 
is numerically costly, and a macroscopic description with viscous fluid 
dynamics would therefore be much preferred if valid. Unfortunately, a 
detailed study presented in \cite{Song:2010aq} indicated that the 
microscopic {\tt UrQMD} dynamics cannot be faithfully simulated with 
viscous hydrodynamics if one assumes the frequently used relationship 
$\tau_\pi T\eq3\frac{\eta}{s}$ between the specific shear viscosity 
$\eta/s$ and the microscopic relaxation time $\tau_\pi$ for the shear 
viscous pressure tensor $\pi^{\mu\nu}$, scaled with the temperature $T$ 
of the fluid. Relations of the type $\tau_\pi T\eq\kappa\frac{\eta}{s}$,
with constant proportionality factors $\kappa$, are found theoretically in
both the extreme weak-coupling (for a massless Boltzmann gas one finds 
$\kappa\eq6$ in Israel-Stewart theory \cite{Israel:1979wp,Baier:2006um} 
and $\kappa\eq5$ in the modified approach by Denicol {\it et al.} 
\cite{Denicol:2010xn}) and extreme strong-coupling limits (where one has 
$\kappa\eq4{-}2\ln2\eq2.614$ for ${\cal N}\eq4$ super-Yang-Mills theory at 
infinite coupling \cite{Bhattacharyya:2008jc,Baier:2007ix,Natsuume:2007ty}).
Other recent work, however, based on modified Kubo relations 
\cite{Koide:2008nw,Koide:2009sy,Moore:2010bu,Huang:2010sa} and a deeper 
analysis of the Boltzmann equation and its connection to viscous 
hydrodynamics \cite{York:2008rr,Betz:2009zz,Denicol:2010xn}, suggests
(in some cases strong) temperature dependence of $\kappa$.

Here we will explore one such proposed relation, $\kappa\eq(e{+}p)/p$ 
\cite{Koide:2009sy} (where $e$ and $p$ are the energy density and pressure 
of the system), which leads to a strong increase of $\kappa(T)$ with 
decreasing temperature in the massive hadron resonance gas below the quark 
confinement temperature $\Tc$. Such an increase is qualitatively
consistent with certain observations made in the recent {\tt VISHNU} study 
\cite{Song:2010aq}. We here use {\tt VISH2{+}1} to investigate, within a 
purely hydrodynamic framework, systematically the consequences of increasing 
shear viscosity and shear pressure relaxation time in the late hadronic stage 
on the transverse momentum spectra and elliptic flow of soft
($p_T{\,<\,}2$\,GeV/$c$) hadrons produced in Au+Au collisions at RHIC.
Our work differs from an earlier study by Bo\.zek \cite{Bozek:2009dw}
of the effects of temperature-dependent specific bulk and shear viscosities
by focusing on shear viscosity and investigating situations in which
the shear viscosity of the hadron gas is {\em larger} than that of
the QGP (rather than the other way around \cite{Bozek:2009dw}), as expected
on basic theoretical grounds \cite{Csernai:2006zz}. While the present work
was being completed, a related study appeared \cite{Niemi:2011ix} which
focusses chiefly on the question whether recent data from Pb+Pb 
collisions at the Large Hadron Collider (LHC) \cite{Aamodt:2010pa}
require an increase of the QGP shear viscosity with rising temperature.

The paper is organized as follows: In Sec.~\ref{sec2} we briefly review
the viscous hydrodynamic model and discuss the specific ingredients used 
in the present study. The effects of a large hadronic specific shear 
viscosity $(\eta/s)_\mathrm{HG}$ on the fireball evolution are discussed
in Sec.~\ref{sec3}. In Sec.~\ref{sec4} we discuss the dependence of the 
transverse momentum spectra and elliptic flow of emitted hadrons in Au+Au 
collisions on $(\eta/s)_\mathrm{HG}$, the decoupling temperature 
$\Tdec$, and the collision centrality. Section~\ref{sec5} is 
dedicated to a detailed discussion of the viscous corrections to the 
freeze-out phase-space distribution and their effects on spectra and 
elliptic flow. All results up to this point assume a constant factor 
$\kappa\eq3$ in the relation $\tau_\pi T\eq\kappa\frac{\eta}{s}$ between 
the specific shear viscosity $\eta/s$ and the microscopic relaxation time 
$\tau_\pi$; in Sec.~\ref{sec6} we explore the consequences of making 
$\kappa(T)$ temperature dependent and letting it grow during the 
quark-hadron phase transition. A final discussion in Sec.~\ref{sec7} 
concludes our paper.

\section{Viscous hydrodynamics: specific ingredients for the present 
study}
\label{sec2}

{\tt VISH2{+}1} \cite{Song:2009gc} solves the second-order Israel-Stewart 
equations for causal relativistic viscous fluid dynamics \cite{Israel:1979wp} 
in the spatial plane transverse to the beam direction and in time, assuming
boost-invariance of the longitudinal expansion. To avoid repetition we 
refer the reader interested in the technical details to earlier descriptions 
of the specific form of the evolution equations and the equation of state
s95p-PCE used here (see specifically Sections II and III in 
Ref.~\cite{Shen:2010uy}). The energy-momentum tensor is decomposed
as $T^{\mu \nu}\eq{e}u^\mu u^\nu - p \Delta^{\mu \nu} + \pi^{\mu \nu}$
where $\pi^{\mu \nu}$ is the viscous pressure tensor. We consider only 
shear viscosity and ignore bulk viscous effects; in this situation
the Israel-Stewart equations describe the evolution of $\pi^{\mu \nu}$
towards it Navier-Stokes limit $2\eta\sigma^{\mu\nu}$ on a microscopic
relaxation time scale $\tau_\pi$, where $\eta$ is the shear viscosity 
and $\sigma^{\mu\nu}$ is the velocity shear tensor which evolves 
hydrodynamically in space and time. We initialize $\pi^{\mu\nu}$ with 
its Navier-Stokes value $\pi^{\mu\nu}\eq2\eta\sigma_{0}^{\mu\nu}$ at 
initial time $\tau_{0}$, calculated from the initial velocity profile 
$u^\mu\eq(u^\tau,u^x,u^y,u^\eta)\eq(1,0,0,0)$.

The generation of hydrodynamic flow from the pressure gradients in the
system is controlled by the fluid's equation of state (EOS) for which
we use s95p-PCE \cite{Huovinen:2009yb,Shen:2010uy} with chemical decoupling
temperature $\Tchem\eq165$\,MeV. This EOS interpolates between 
state-of-the-art Lattice QCD data at high temperatures and a chemically 
frozen hadron resonance gas at low temperatures. Chemical freeze-out at
$\Tchem\eq165$\,MeV guarantees that the final hadron yields, calculated
by integrating the final hadron momentum spectra obtained from the 
hydrodynamic output along an isothermal decoupling surface of temperature 
$\Tdec{\,<\,}\Tchem$ via the Cooper-Frye procedure \cite{Cooper:1974mv}
followed by resonance decay \cite{Sollfrank:1990qz,Sollfrank:1991xm},
agree with experimental measurements in 200\,$A$\,GeV Au+Au collisions
at RHIC \cite{Adams:2005dq,BraunMunzinger:2001ip,Andronic:2008gu}.
At decoupling, we parametrize the local distribution function in
the Cooper-Frye formula by a local thermal equilibrium function plus a
small viscous correction which depends on the value of the viscous
pressure tensor $\pi^{\mu\nu}$ on the freeze-out surface and increases
quadratically with particle momentum \cite{Teaney:2003kp,Shen:2010uy}.
Unless noted otherwise, we use $\Tdec\eq120$\,MeV.

We initialize the hydrodynamic evolution with an energy density profile 
obtained from the optical {\tt fKLN} model \cite{Drescher:2006pi,%
Drescher:2006ca,fKLN}. The model yields the initial gluon density
distribution which, after thermalization, gives directly the initial
entropy density which is then converted to energy density using the EOS 
s95p-PCE. The normalization of the initial entropy density is adjusted
in the most central collisions to reproduce the finally measured charged 
hadron multiplicity. Due to viscous entropy production, changing $\eta/s$ 
requires a readjustment of this normalization to keep the final
multiplicity fixed. After normalization in central collisions, the 
centrality dependence of the final charged hadron multiplicity is
obtained directly from the {\tt fKLN} model, without further adjustment 
of parameters. 

The key ingredients whose influence on the generation of radial
and elliptic flow we want to study here are the temperature dependence 
of the specific shear viscosity $\eta/s$ and of the proportionality 
constant between $\eta/s$ and the temperature-scaled microscopic 
relaxation time $\tau_\pi T$, $\kappa\eq\frac{\tau_\pi T}{\eta/s}$.
Specifically, we will explore scenarios where $\eta/s\eq0.16$ is a 
constant in the QGP phase but increases by variable amounts during the 
transition from QGP to hadrons, using the following parametrization for 
its temperature dependence:
\begin{eqnarray}
\notag
  \frac{\eta}{s}(T) &=& 
  \frac{(\eta/s)_{\mathrm{QGP}}+(\eta/s)_{\mathrm{HG}}}{2} 
\\
  &+& \frac{(\eta/s)_{\mathrm{QGP}}-(\eta/s)_{\mathrm{HG}}}{2} 
  \tanh\left(40\frac{T{-}\Tc}{\Tc}\right).
\label{eq1}
\end{eqnarray}
%
\begin{figure}[h!]
\includegraphics[width=0.95\linewidth,clip=]{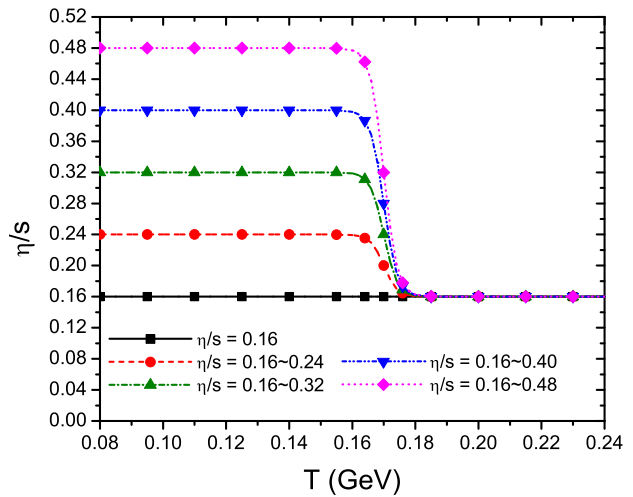}
\caption{\label{F1} (Color online)
Five choices for the temperature dependent $(\eta/s)(T)$ studied in this 
work. The constant values at low $T$ are multiples of 
$0.08\approx\frac{1}{4\pi}$.
}
\end{figure}
%
%
Here $\Tc\eq170$\,MeV, and $(\eta/s)_\mathrm{QGP}\eq0.16$ 
and $(\eta/s)_\mathrm{HG}$ are (different) constants for the 
QGP and HG (hadron gas) phases. We will explore the range 
$0.16{\,\leq\,}(\eta/s)_\mathrm{HG}$ ${\,\leq\,}0.48$, as illustrated
in Figure~\ref{F1}. In the next three sections $\kappa$ will be held
constant at $\kappa\eq3$;\footnote{The specific values 
   $(\eta/s)_{\mathrm{QGP}}\eq0.16$ and $\kappa\eq3$ chosen here 
   agree with those used by us in the earlier studies 
   \cite{Shen:2010uy,Song:2010mg} while the recent work 
   \cite{Niemi:2011ix} assumes $\kappa\eq5$.}
consequences of a temperature dependent $\kappa(T)\eq(e{+}p)/p$ will be 
explored in Sec.~\ref{sec6}.

\section{Hydrodynamic evolution}
\label{sec3}

In order to study how the fireball evolves with a temperature 
dependent $(\eta/s)(T)$ that increases in the HG phase, we graph the 
time evolution for the average transverse flow velocity $\lda v_\perp\rda$ 
(the average over the transverse plane being defined with the lab-frame 
energy density $\gamma_{\perp} e$ as weight), the spatial eccentricity 
$\ecc_x\eq\frac{\lda y^2{-}x^2 \rda}{\lda y^2{+}x^2 \rda}$
of the lab-frame energy density distribution, the {\em flow} momentum 
anisotropy $\ecc_p\eq\frac{\langle T_0^{xx}{-}T_0^{yy} \rangle}
{\langle T_0^{xx}{+}T_0^{yy} \rangle}$ (where $\La\dots\Ra$ denotes simple
integration over the transverse plane and  $T_0^{\mu \nu}$ is the
ideal fluid part of the energy-momentum tensor, without viscous
pressure contributions), and the {\em total} momentum anisotropy
$\ecc'_p\eq\frac{\langle T^{xx}{-}T^{yy}\rangle}{\langle 
T^{xx}{+}T^{yy}\rangle}$ for different choices of the temperature
dependence of $\eta/s$. 

%
\begin{figure}[hbt]
\includegraphics[width=0.96\linewidth,clip=]{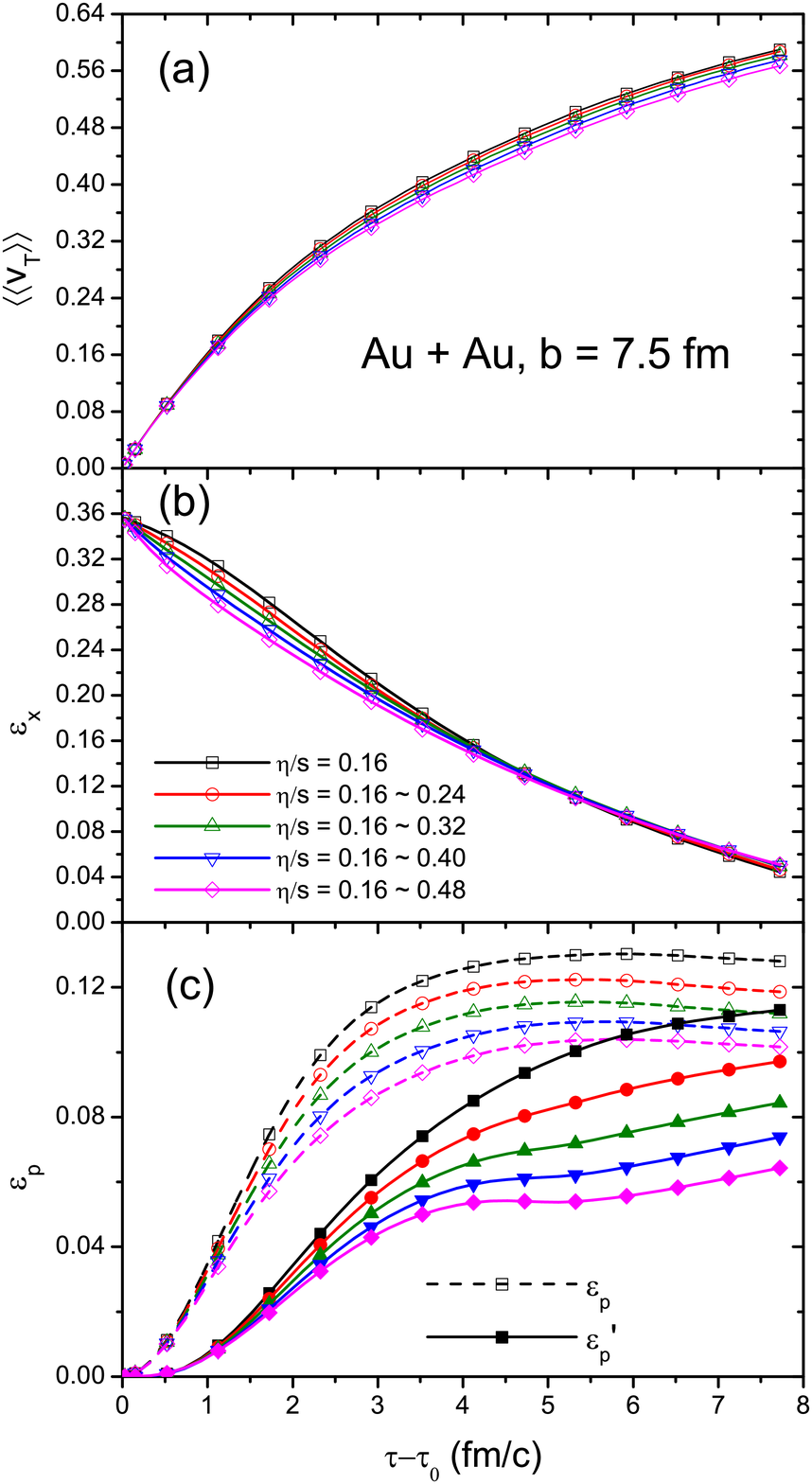}
\caption{\label{F2} (Color online)
The average radial flow $\lda v_\perp \rda$, spatial eccentricity 
$\ecc_x$, and the flow and total momentum anisotropies $\ecc_p$
and $\ecc'_p$ for Au+Au collisions at $b\eq7.5$\,fm as functions 
of hydrodynamic evolution time $\tau{-}\tau_0$, for $\tau_0\eq0.4$\,fm/$c$
and kinetic freeze-out temperature $\Tdec\eq120$\,MeV. Lines with different 
symbols correspond to different temperature dependences of $\eta/s$ as
shown in Fig.~\ref{F1}.
}
\end{figure}
%

Since shear viscosity leads to viscous heating which generates entropy, 
holding the finally observed hadron multiplicity fixed requires that
an increase in $(\eta/s)(T)$ must be accompanied by a decrease of the 
initial entropy of the fireball. We implement this by a decrease of the
normalization of the initial entropy density distribution, keeping its 
shape fixed. Whereas for fixed initial conditions an overall increase of 
$\eta/s$ leads to stronger radial acceleration due to a positive 
contribution from the viscous pressure tensor $\pi^{\mu\nu}$ to the 
transverse pressure gradients \cite{Teaney:2003kp,Chaudhuri:2005ea,%
Baier:2006gy,Song:2007fn,Song:2007ux}, this effect is largely compensated 
\cite{Romatschke:2007jx,Shen:2010uy,Niemi:2011ix} after rescaling the 
initial entropy density to ensure fixed final multiplicity. For our 
temperature-dependent $\eta/s$ this compensation no longer works in the
same way: after rescaling the initial entropy density profile, to compensate 
for increased viscous heating in the hadronic phase, the QGP core shrinks 
and the HG corona grows in size. Since the viscous pressure is relatively 
larger in the hadronic phase than in the QGP, the {\em effective} transverse 
pressure gradient is {\em reduced} when increasing $\eta/s$ only in the 
hadronic phase, leading to weaker radial acceleration. This can be seen in 
Fig.~\ref{F1}a, where we see a reduction oof the growth rate of the average 
radial flow velocity $\lda v_\perp\rda$ with increasing values of 
$(\eta/s)_\mathrm{HG}$, holding $(\eta/s)_\mathrm{QGP}\eq0.16$ fixed. 
 
%
\begin{figure*}[t!]
\includegraphics[width=\linewidth,clip=]{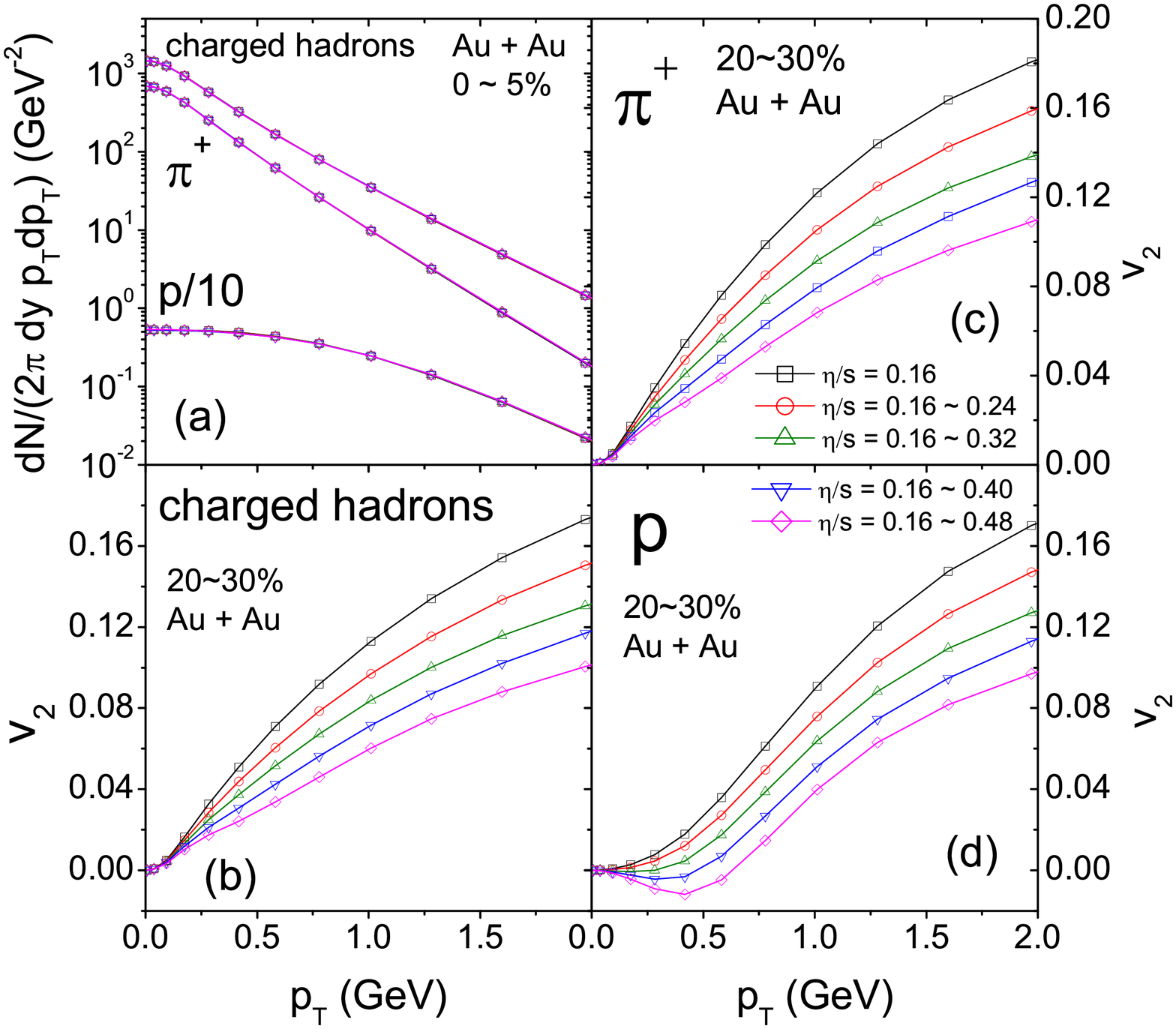}
\caption{\label{F3} (Color online)
(a): Transverse momentum spectra for charged hadrons, pions, and protons 
from {\tt VISH2+1} for the 5\% most central Au+Au collisions ($b\eq2.33$\,fm). 
(b-c): Differential elliptic flow $v_2(p_T)$ for charged hadrons (b), pions 
(c) and protons (d) from Au+Au collisions at 20-30\% centrality 
($b\eq7.49$\,fm). Lines with different symbols correspond to
different values of $(\eta/s)_\mathrm{HG}$ as shown in Fig.~\ref{F1};
$\Tdec\eq120$\,MeV. Decay products from all strong resonance decays are 
included. Charged hadrons include $\pi^+$, $K^+$, $p$, $\Sigma^\pm$, 
$\Xi^{-}$, $\Omega^{-}$, and their antiparticles.
}
\end{figure*}
%

The larger shear viscosity in the hadronic corona leads to a more rapid
initial decay\footnote{We note that $\ecc$ is defined by integrating
  at fixed time $\tau$ over the entire transverse plane, including both
  thermalized and already decoupled matter. It is possible that the strong
  initial decay of $\ecc$ seen in Fig.~\ref{F2}b arises mostly from
  contributions in that part of the hadronic corona that has already
  decoupled.}
of the spatial fireball eccentricity $\ecc_x$ (see Fig.~\ref{F2}b) and a 
slower growth rate and lower asymptotic value of 
the flow momentum anisotropy $\ecc_p$ (Fig.~\ref{F2}c, open symbols).
The spatial eccentricity curves in Fig.~\ref{F2}b all cross around
$\tau{-}\tau_0\eq4.5$\,fm/$c$, indicating the transition from stronger 
decay of $\ecc_x$ at early times to weaker decay at late times for
larger values of $(\eta/s)_\mathrm{HG}$. This is a consequence of the
reduced flow anisotropy $\ecc_p$ shown in Fig.~\ref{F2}c.
 
The lines with filled symbols in Figure~\ref{F2}c show that the effects
of increased hadronic viscosity on the asymptotic values of the total 
momentum anisotropy $\ecc'_p$ are much stronger than on the flow 
anisotropy $\ecc_p$: while the latter decreases by about 25\% from 
$(\eta/s)_\mathrm{HG}\eq0.16$ to $(\eta/s)_\mathrm{HG}\eq0.48$, the 
corresponding decrease for $\ecc'_p$ is almost twice as large. Also,
most of the effect on $\ecc'_p$ happens at late times 
$\tau{-}\tau_0{\,>\,}4.5$\,fm/$c$ when most of the matter has converted
into hadron gas. This reflects the growth of the Navier-Stokes value
$\pi^{\mu\nu}_\mathrm{NS}\eq2\eta\sigma^{\mu\nu}$ of the viscous 
pressure contribution to $T^{\mu\nu}$ in the hadronic phase where 
$\eta_\mathrm{HG}$ increases. In contrast to $\ecc_p$, the total
momentum anisotropy $\ecc'_p$ does not saturate at late times after 
the spatial eccentricity (which drives the flow anisotropy) has
essentially decayed to zero; its continued increase is due to
the continuing decrease of the magnitude of the $\pi^{\mu\nu}$ 
components whose contribution to $\ecc'_p$ is negative \cite{Song:2007ux}.

%
\begin{figure*}
\includegraphics[width=0.95\linewidth,clip=]{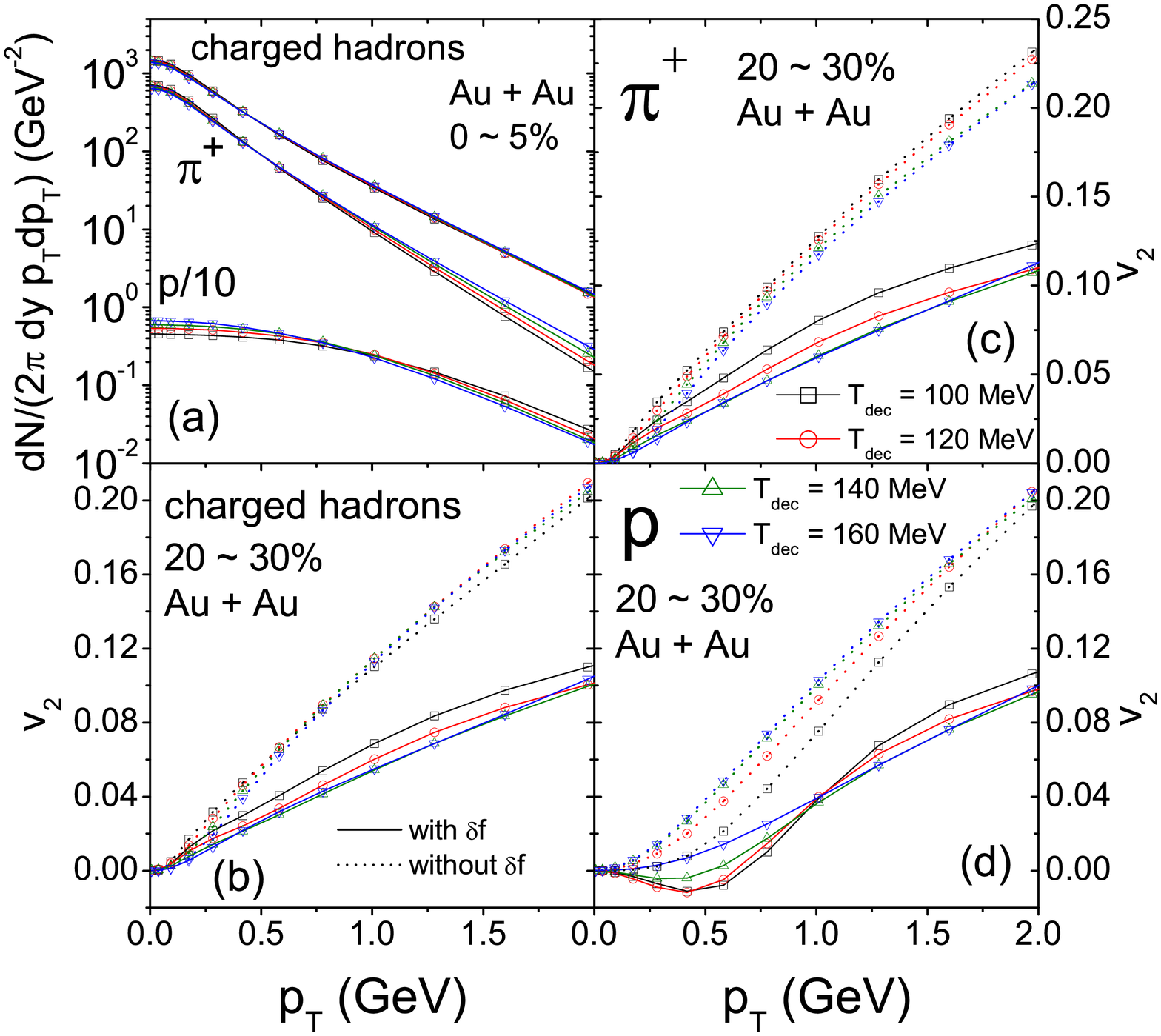}
\caption{\label{F4} (Color online)
Similar to Fig.~\ref{F3}, for fixed $(\eta/s)_\mathrm{HG}\eq0.48$ and
different decoupling temperatures $\Tdec$ ranging from 160 to 100\,MeV.
In panels (b)-(d), dotted lines show $v_2(p_T)$ calculated without
the $\delta f$ correction whereas the solid lines show the full calculations.
}
\end{figure*}
%

The large difference between the late-time values of $\ecc_p$ and
$\ecc'_p$ for high values of $\eta_\mathrm{HG}$ shows that, for strong
hadronic viscosity, the viscous corrections to the local thermal equilibrium
distribution on the kinetic decoupling surface at $\Tdec$ are big. We will
explore this in more detail in Sec.~\ref{sec5}.

\section{Spectra and elliptic flow}
\label{sec4}
\subsection{Central and semi-peripheral Au+Au collisions}
\label{sec4a}

Figure~\ref{F3} shows the transverse momentum spectra for charged hadrons,
pions and protons from central Au+Au collisions ($0{-}5\%$ centrality) and 
their elliptic flows $v_2(p_T)$ for semiperipheral Au+Au collisions 
($20{-}30\%$ centrality) for different choices of the hadronic shear 
viscosity $(\eta/s)_\mathrm{HG}$.
The $p_T$-spectra in Fig.~\ref{F3}a are seen to be completely insensitive
to the value of $(\eta/s)_\mathrm{HG}$. From the reduction of the radial
flow seen in Fig.~\ref{F2}a one would have expected steeper spectra for
larger $(\eta/s)_\mathrm{HG}$ since $\Tdec\eq120$\,MeV is held fixed;
clearly, for $p_T<2$\,GeV/$c$, the viscous correction $\delta f$ to the 
local equilibrium distribution at freeze-out (which will be analyzed
in greater depth in Sec.~\ref{sec5}) happens to almost exactly 
compensate for the loss of radial flow, over the entire range of 
$(\eta/s)_\mathrm{HG}$ values studied here. 

This is not true for the elliptic flow which is strongly reduced when
the hadronic viscosity is increased (Figs.~\ref{F3}b-d). For protons a 
striking effect is seen for $(\eta/s)_\mathrm{HG}{\,>\,}0.32$: The 
proton elliptic flow turns {\em negative} (i.e. protons show stronger 
flow perpendicular than parallel to the reaction plane) for low $p_T$.
This effect is caused entirely by the $\delta f$ correction. $\delta f$ 
grows not only with $p_T$, as is well known, but also with the mass of the 
hadron. For massive hadrons, the shear viscous $\delta f$ correction can 
be a strong effect even at $p_T\eq0$. In Fig.~\ref{F3} negative $v_2(p_T)$ 
caused by $\delta f$ at low $p_T$ is not visible for pions, but for protons
and would be much stronger for $\Omega$ hyperons or $J/\psi$ mesons 
\cite{Chun_Jpsi} if they also followed viscous hydrodynamical evolution 
down to $\Tdec\eq120$\,MeV.

The effect of the $\delta f$ correction is studied in Fig.~\ref{F4}, for
various choices of the decoupling temperature $\Tdec$ . We hold the 
hadronic shear viscosity fixed at $(\eta/s)_\mathrm{HG}\eq0.48$,
the largest value studied here. The effect of variations in $\Tdec$ on 
the spectra in Fig.~\ref{F4}a is similar to what we observed in 
\cite{Shen:2010uy}: lower decoupling temperatures cause flatter proton 
spectra due to larger radial flow, steeper pion spectra due to the cooling 
effect which dominates for light particles, and almost no change in the 
charged hadron spectra whose mix of light and heavy particles effectively 
balances the counteracting cooling and radial flow effects.

In Figs.~\ref{F4}b-d we plot the differential elliptic flow for charged
hadrons, pions and protons. The dotted lines show a calculation that 
ignores the viscous $\delta f$ correction at freeze-out and thus only
includes the $\Tdec$-dependence of the pure flow effects. We see that
lower $\Tdec$ values suppress $v_2(p_T)$ for protons but increase it for 
pions at low $p_T$. This is really a consequence of the accompanying 
change of the $p_T$-spectra: Due to the large hadronic viscosity, very 
little additional flow momentum anisotropy is generated at temperatures 
below $\Tc$. However, due to cooling, the pion spectra get steeper with
decreasing $\Tdec$, moving more of their momentum anisotropy to low
transverse momenta which leads to the increase of pion $v_2(p_T)$ at low 
$p_T$. Conversely, the proton spectra get flatter, in spite of cooling, 
due to additional radial flow developing between $\Tc$ and $\Tdec$;
consequently, their total momentum anisotropy gets shifted on average
to larger transverse momenta, causing a reduction of proton $v_2(p_T)$
at low $p_T$ (accompanied by an increase at high $p_T<2$\,GeV/$c$,
beyond the range shown here). Both the flattening of the proton spectra
and the shifting of their elliptic flow to larger $p_T$ are stronger
for the case of large hadronic shear viscosity ($(\eta/s)_\mathrm{HG}\eq0.48$)
studied here than for the case of temperature-independent $\eta/s\eq0.16$
studied in \cite{Shen:2010uy}: The large hadronic viscosity generates
stronger additional radial flow but less additional momentum anisotropy
in the hadronic stage than does constant $\eta/s\eq0.16$. Note that, 
without $\delta f$, proton $v_2(p_T)$ never turns negative, even for the 
largest hadronic shear viscosity studied in this work.

The solid lines in Figs.~\ref{F4}b-d show the full calculation of 
$v_2(p_T)$ including the $\delta f$ correction. We see larger
$\delta f$ effects for protons than pions, due to their larger rest 
mass \cite{Song:2007ux}. The full calculations feature a non-monotonic
variation of pion and charged hadron $v_2(p_T)$ with decoupling 
temperature $\Tdec$: The suppression from $\delta f$ is smaller for 
$\Tdec\eq160$ MeV than for $\Tdec\eq140$\,MeV. The like explanation 
is that $\Tdec\eq160$\,MeV is so close to the inflection point $\Tc$ of 
the shear viscosity $(\eta/s)(T)$ that, due to the finite relaxation time
$\tau_\pi\sim2$\,fm/$c$ at this temperature, the viscous pressure
tensor has not yet had time to fully evolve to its (larger) hadronic
Navier-Stokes value whereas at $\Tdec$ complete relaxation has been
achieved. At sufficiently low $\Tdec$, $\delta f$ decreases with
decreasing the decoupling temperature, since now $\eta/s$ has reached
its new, higher hadronic level and $\pi^{\mu\nu}$ becomes smaller simply 
due to hydrodynamic expansion \cite{Song:2007ux}.

\subsection{Minimum bias collisions}
\label{sec4b}

\begin{figure}[b!]
\includegraphics[width=0.9\linewidth,clip=]{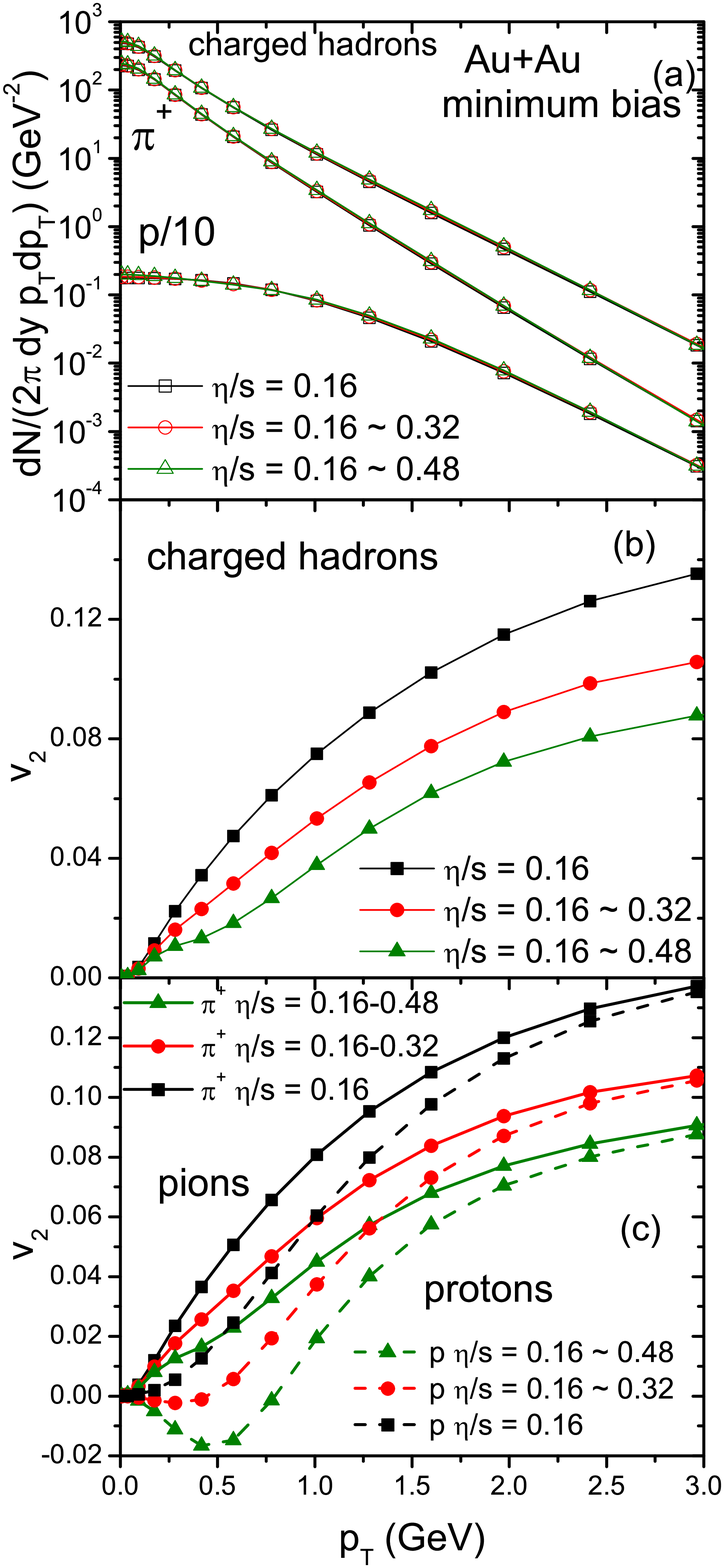}
\caption{\label{F5} (Color online)
Transverse momentum spectra (a) and differential elliptic flow
$v_2(p_T)$ for charged hadrons (b) and pions and protons (c) from
minimum bias Au+Au collisions at RHIC, for various $(\eta/s)(T)$ as 
indicated ({\em c.\,f.} Fig.~\ref{F1}).
\vspace{-5mm}
}
\end{figure}
%

In Figure~\ref{F5} we show $p_T$ spectra and differential elliptic flow
for charged hadrons, pions and protons from minimum bias Au+Au collisions
with $\Tdec\eq120$\,MeV. For these we summed our calculated results over 
all collision centralities ${\leq\,}80\%$. The dependence on collision 
centrality is discussed in the next subsection. 

Similar to what we saw in Fig.~\ref{F3}a for central Au+Au collisions,
the spectra shown in Fig.~\ref{F5}a exhibit almost no sensitivity at all
to variations of the specific shear viscosity $(\eta/s)_\mathrm{HG}$ in
the hadron gas stage. We did observe some flattening of the charged
hadron spectrum in the most peripheral ($70{-}80\%$) centrality bin 
studied, where the viscous effects are strongest and the $\delta f$ 
correction is largest. Due to its low weight in the average, this weak
effect is not visible in the minimum bias result.

In Figures~\ref{F5}b and c, the minimum bias differential $v_2(p_T)$ of all 
charged hadrons, pions and protons are shown for different 
$(\eta/s)_\mathrm{HG}$. We see that the features observed in Fig.~\ref{F3}
for the specific $20{-}30\%$ centrality bin carry over, qualitatively 
unchanged, to event samples without centrality selection: a significant 
increase of $\eta/s$ in the hadron gas phase has a strong suppression 
effect on $v_2(p_T)$. However, as shown in Sec.~\ref{sec4a}, the suppression 
arises mostly from the $\delta f$ correction at kinetic freeze-out, with a 
much smaller contribution accounting for the lack of growth of the total 
momentum anisotropy in the hadronic phase when $(\eta/s)_\mathrm{HG}$ 
becomes large. Hence, the strong suppression of differential elliptic flow 
by large hadronic shear viscosity shown here is critically dependent on the 
validity of viscous hydrodynamics as the correct framework for evolving 
$\delta f$ all the way down to $\Tdec\eq120$\,MeV. This is assumed here, 
but not supported by the analysis presented in \cite{Song:2010aq}.

\subsection{Centrality dependence of elliptic flow}
\label{sec4c}

\begin{figure}[t!]
\includegraphics[width=0.95\linewidth,clip=]{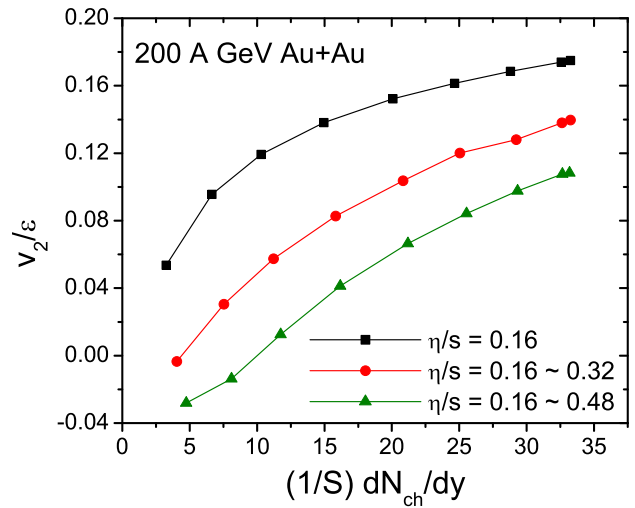}
\caption{\label{F6} (Color online)
Eccentricity-scaled charged hadron elliptic flow $v_2/\ecc$ as a 
function of the multiplicity density $(1/S)(\dNdy)$, for different values of 
$(\eta/s)_{HG}$. The overlap area $S\eq\pi\sqrt{\lda x^2\rda\lda y^2\rda}$ is 
calculated from the same initial profiles as the spatial eccentricity 
$\ecc$. 
}
\end{figure}

The centrality dependence of the eccentricity-scaled elliptic flow 
$v_2/\ecc$ is shown in Fig.~\ref{F6} where we graph this quantity
%
%
as a function of the final charged multiplicity density $(1/S)\dNdy$.
for different values of $(\eta/s)_\mathrm{HG}$. (We obtain $v_2$ by
integrating $v_2(p_T)$ over all $p_T$, without regard to possible $p_T$
cuts imposed by experimental constraints.) Strong suppression of $v_2/\ecc$ 
by hadronic viscosity is observed even in the most central collisions, but 
the effect is stronger in peripheral collisions. An increase of 
$(\eta/s)_\mathrm{HG}$ thus not only decreases $v_2/\ecc$, but also
changes the slope of its centrality dependence. We note in passing that 
in recent studies with the hydro+cascade hybrid code \VC\ \cite{Song:2010mg} 
this slope was fixed and controlled by the effective dissipation encoded 
in the hadron cascade, and that in \cite{Song:2010aq} an (unsuccessful) 
attempt was made to extract the temperature-dependence of 
$(\eta/s)_\mathrm{HG}$ (here assumed to be $T$-independent) by matching the 
magnitude and slope of the corresponding $v_2/\ecc$ vs. $(1/S)\dNdy$ curves 
from \VISH\ to those from \VC. We also observe that for the largest value
of $(\eta/s)_\mathrm{HG}$ studied here, $(\eta/s)_\mathrm{HG}\eq0.48$,
the total charged hadron elliptic flow turns negative in the most peripheral
($70{-}80\%$) centrality bin. We found that this is caused by negative
{\em pion} $v_2(p_T)$ around $p_T\eq0.5$\,GeV/$c$ (i.e. close to
their average $p_T$), caused by large $\delta f$ corrections at 
freeze-out.\footnote{For $70{-}80\%$ centrality and 
  $(\eta/s)_\mathrm{HG}\eq0.32{-}0.48$, we found for that for pions
  $v_2(p_T)$ first rises at very low $p_T$, then turns negative for
  $0.25{\,<\,}p_T{\,<\,}0.75$\,GeV/$c$ before turning positive again
  and continuing to grow approximately linearly with $p_T$. This is
  different from protons whose $v_2(p_T)$ turns negative right away
  at small $p_T$, again with a minimum around 0.5\,GeV/$c$. All these
  effects are caused by large $\delta f$ effects; in this centrality
  bin we do not trust viscous hydrodynamic predictions to be very robust.}

\section{$\bm{\delta f}$ contributions}
\label{sec5}

%
\begin{figure*}
\includegraphics[width=1.0\linewidth,clip=]{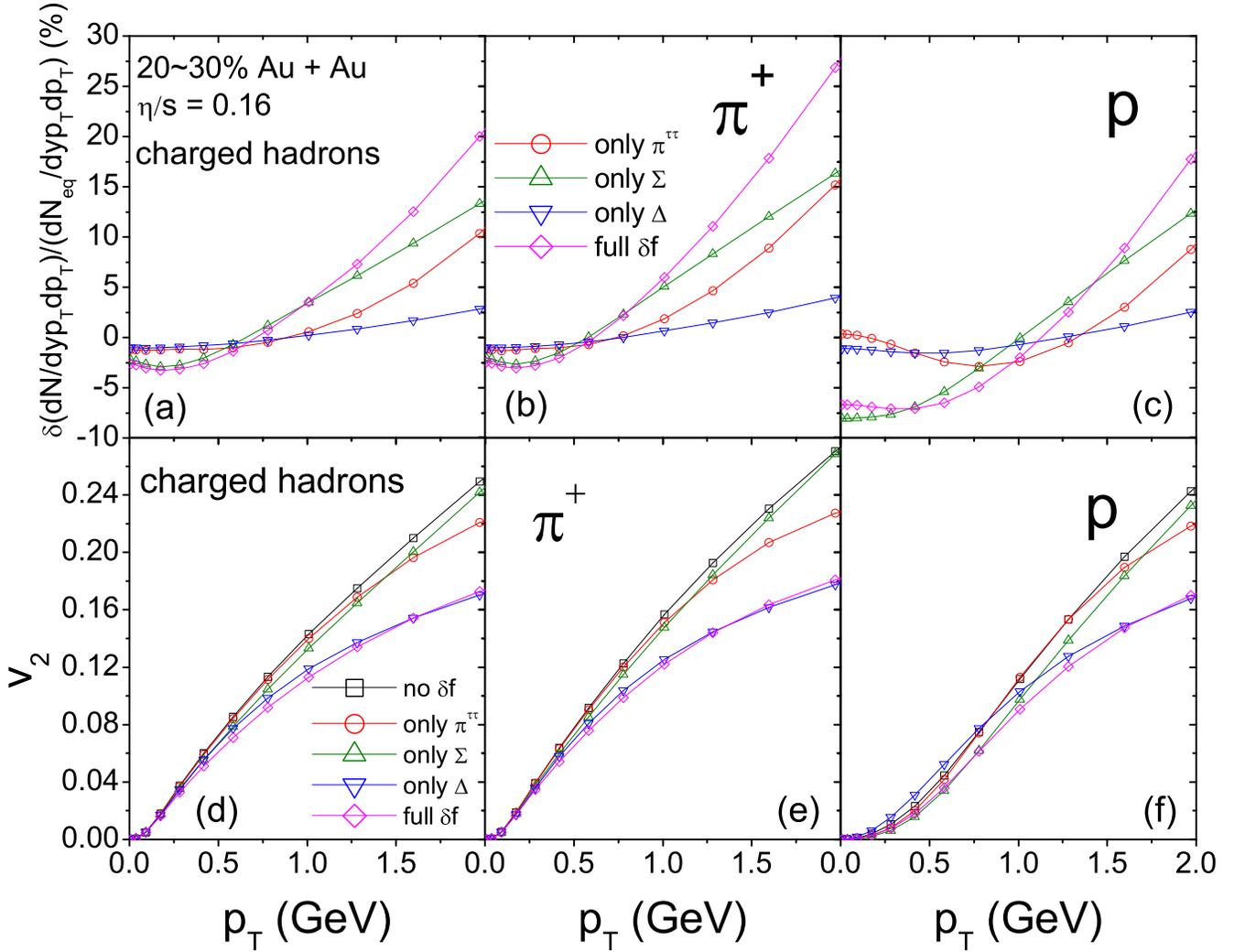}
\caption{\label{F7} (Color online)
The $\delta f$ correction for constant $\eta/s\eq0.16$ to the 
$p_T$-spectra for charged hadrons (a), pions (b) and protons (c)
at $0{-}5\%$ centrality, and to their differential elliptic flow (d-f)
at $20{-}30\%$ centrality. Lines with different symbols denote individual 
contributions as described in the text.
}
\end{figure*}
%

Due to non-zero viscous pressure components $\pi^{\mu\nu}$, the
distribution function $f_i(x,p)$ for hadron species $i$ must deviate 
on the freeze-out surface from local equilibrium:
\begin{equation}
  f_{i} (x,p) = f_{\mathrm{eq},i}(x,p) + \delta f_{i}(x,p).
  \label{eq3}
\end{equation}
We use \cite{Teaney:2003kp} 
\begin{equation}
  \label{eq4}
  \delta f_{i} = f_{\mathrm{eq},i} \cdot \frac{1}{2} 
  \frac{p^{\mu}p^{\nu}}{T^2}\frac{\pi_{\mu\nu}}{e+p},
\end{equation}
noting that also other forms have been suggested in the literature 
\cite{Monnai:2009ad,Dusling:2009df}. The numerator can be written as
\begin{eqnarray}
  && \!\!\!\!\!\!
     p^{\mu}p^{\nu}\pi_{\mu\nu}(x)
\nonumber\\ 
  && = \pi^{\tau\tau}(x) \biggl[
     m^2_T \bigl(2 \cosh^2(y{-}\eta)-1\bigr) 
     + \frac{p^2_T}{v^2_\perp} \frac{\sin(2\phi_p)}{\sin(2\phi_v)}
\nonumber\\ 
  &&\qquad\qquad\
     - 2\frac{p_T}{v_\perp} m_T\cosh(y{-}\eta) 
        \frac{\sin(\phi_p{+}\phi_v)}{\sin(2\phi_v)} 
     \biggr]
\nonumber\\
  &&\ \ +\ \Sigma(x) \biggl[
      - m^2_T \sinh^2(y{-}\eta) 
      + \frac{p_T^2}{2}\Bigl(1 - \frac{\sin(2\phi_p)}{\sin(2\phi_v)}\Bigr) 
      \biggr.
\nonumber\\
  &&\qquad\qquad\
      \biggl.
      + p_T m_T \cosh(y{-}\eta) v_\perp 
       \frac{\sin(\phi_p{-}\phi_v)}{\tan(2\phi_v)} 
     \biggr]
\nonumber\\ 
  &&\ \ +\ \Delta(x) \biggl[
      p_T m_T \cosh(y{-}\eta) v_{\perp} 
       \frac{\sin(\phi_p{-}\phi_v)}{\sin(2\phi_v)}
      \biggr.
\nonumber\\
  &&\qquad\qquad\
     \biggl.
     - \frac{p_T^2}{2} \frac{\sin(2(\phi_p{-}\phi_v))}{\sin(2\phi_v)}
     \biggr] 
\label{eq5}
\end{eqnarray}
where $\Sigma\eq\pi^{xx}{+}\pi^{yy}$, $\Delta\eq\pi^{xx}{-}\pi^{yy}$.
Because of boost-invariance, tracelessness and orthogonality to $u^{\mu}$, 
only three components of $\pi^{\mu\nu}$ are independent; we take them as 
$\Sigma$, $\Delta$, and $\pi^{\tau\tau}$. $m_T\eq\sqrt{m^2 + p_T^2}$ 
is the transverse mass of the particles, $\phi_p$ is the azimuthal angle 
of $\bm{p}_T$, and $\phi_v(x)$ is the azimuthal angle of the fluid velocity 
$\bm{v}$ at point $x$. 

%
\begin{figure*}
\includegraphics[width=1.0\linewidth,clip=]{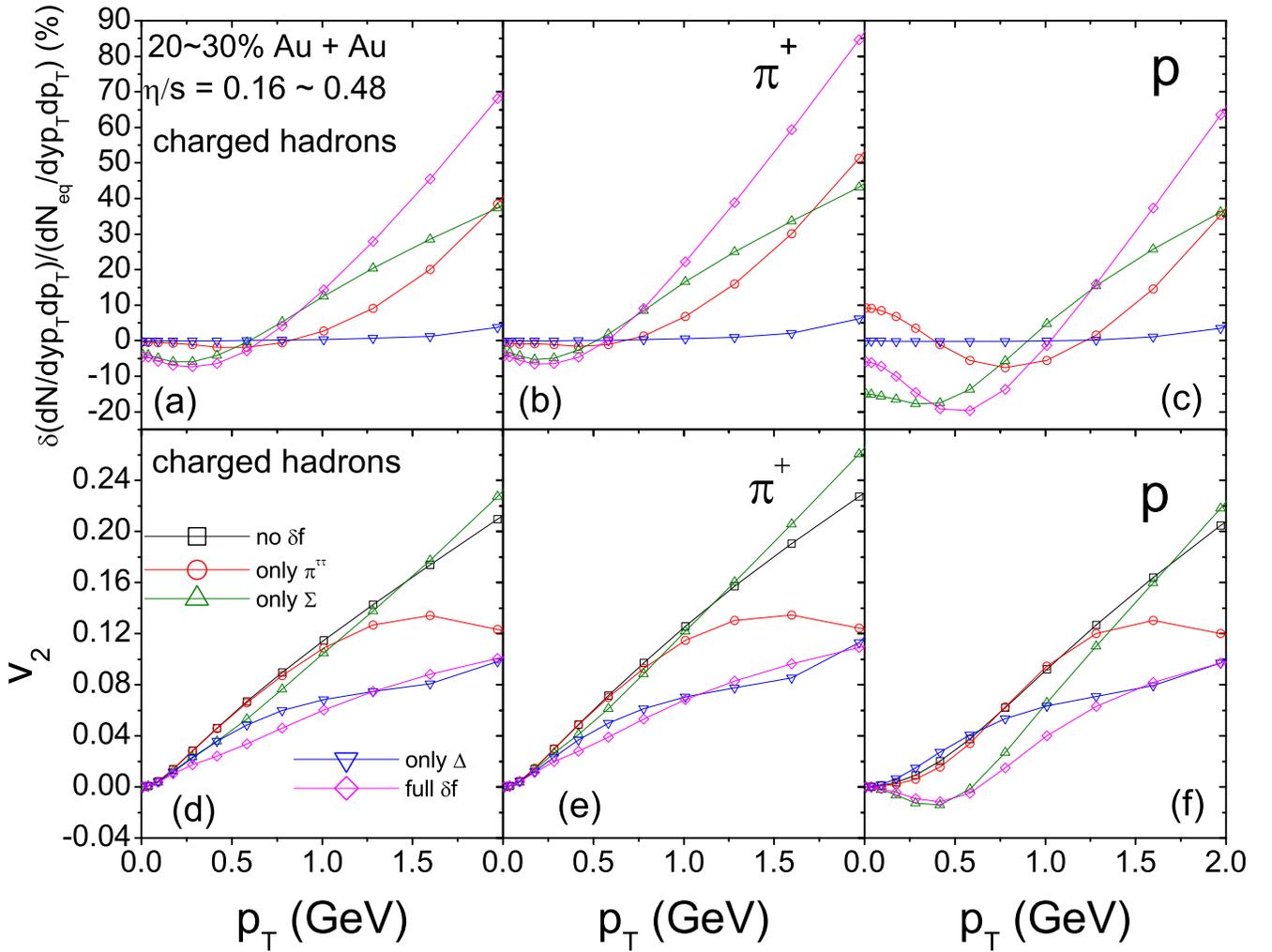}
\caption{\label{F8} (Color online)
Similar to Fig.~\ref{F7}, but for temperature dependent $(\eta/s)(T)$,
Eq.~(\ref{eq1}), with $(\eta/s)_\mathrm{HG}\eq0.48$.
}
\end{figure*}
%

We now discuss the individual contributions from Eq.~(\ref{eq5}) to
the $p_T$-spectra and elliptic flow, for the cases of constant 
$\eta/s\eq0.16$ (Figure~\ref{F7}) and temperature-dependent $(\eta/s)(T)$
(Figure~\ref{F8}). In panels (a-c) we show the fractional contribution 
$\delta N/N_\mathrm{eq}$ from $\delta f$ to the Cooper-Frye spectra of 
charged hadrons (a), pions (b) and protons (c). At low $p_T$, the 
contributions proportional to $\pi^{\tau\tau}$ and $\Delta$ (first and 
last terms on the r.h.s. of Eq.~(\ref{eq5})) are small and overshadowed 
by the contribution from the average transverse viscous pressure $\Sigma$. 
The first (negative) term $\sim {-}m_T^2$ in the expression multiplying 
$\Sigma$ dominates at low $p_T$. It obviously grows with rest mass, leading 
to large negative $\delta N/N_\mathrm{eq}$ corrections at low $p_T$ for 
heavy hadrons such as $\Omega$ and $J/\psi$ \cite{Chun_Jpsi}. For
protons the effect remains below 10\% in central Au+Au collisions, i.e.
$\delta f$ corrections are small and the calculation is reliable. At
larger $p_T$, all three contributions in Eq.~(\ref{eq5}) turn positive
and $\delta N/N_\mathrm{eq}$ switches sign (around 0.5\,GeV/$c$ for pions
and around 1\,GeV/$c$ for protons). Again, the term ${\sim\,}\Sigma$
first dominates, but since it grows only linearly at large $p_T$ it is
eventually (at $p_T{\,\agt\,} 2$\,GeV/$c$) overtaken by the term 
${\sim\,}\pi^{\tau\tau}$. For constant $\eta/s\eq0.16$, 
$|\delta N/N_\mathrm{eq}|$ remains below 25\% up to $p_T\eq2$\,GeV/$c$ 
for all three spectra shown,\footnote{The $\delta f$ effects on charged 
  hadron spectra can be qualitatively understood from those on pion
  and proton spectra by noting that at low $p_T$ charged hadrons are 
  dominated by pions whereas at larger $p_T$ heavier hadrons become 
  increasingly more important.} 
and the calculation is therefore reliable.
For large hadronic viscosity $(\eta/s)_\mathrm{HG}\eq0.48$ (Fig.~\ref{F8})
the $\delta f$ corrections to the $p_T$-spectra are larger, in particular
the term ${\sim\,}\pi^{\tau\tau}$, and $|\delta N/N_\mathrm{eq}|$ reaches
$70{-}80\%$ at $p_T\eq2$\,GeV/$c$, indicating the imminent breakdown
of the viscous hydrodynamic expansion $|\delta f|{\,\ll\,}f_\mathrm{eq}$.

In the lower panels of Figs.~\ref{F7} and \ref{F8} we show the $\delta f$
contributions to the differential $v_2(p_T)$ for charged 
hadrons (d), pions (e), and protons (f), again separated into their
individual contributions according to Eq.~(\ref{F3}). We see that for
low $p_T$ all three terms in Eq.~(\ref{eq5}) contribute to the suppression 
of elliptic flow, but that in this case at high $p_T$ the term proportional 
to the viscous pressure {\em anisotropy} $\Delta\eq\pi^{xx}{-}\pi^{yy}$  
plays the dominant role, overshadowing the terms ${\sim\,}\Sigma$ and 
(except for the largest hadronic viscosities) also ${\sim\,}\pi^{\tau\tau}$. 
The latter grows quadratically with $p_T$ and eventually wins over the term 
${\sim\,}\Sigma$; for large hadronic viscosity (Fig.~\ref{F8}) it even 
exceeds the anisotropy term ${\sim\,}\Delta$ at sufficiently large $p_T$. 
The term proportional to the average transverse viscous pressure $\Sigma$ 
individually generates a positive elliptic flow correction at large $p_T$ 
(i.e. at $p_T{\,\agt\,}2$\,GeV/$c$ for constant $\eta/s\eq0.16$
and at $p_T{\,\agt\,}(1{-}1.5)$\,GeV/$c$ for $T$-dependent $(\eta/s)(T)$ with 
$(\eta/s)_\mathrm{HG}\eq0.48$). Similarly the anisotropy term 
${\sim\,}\Delta$ by itself increases proton elliptic flow at low $p_T$ 
if the hadronic viscosity is large enough (Fig.~\ref{F8}f). In the sum,
however, these positive individual corrections are always overwhelmed by 
the remaining two negative corrections, leading to an overall suppression 
of $v_2(p_T)$ at all $p_T$ in all cases. Interestingly, the negative proton 
elliptic flow at low $p_T$ and large $(\eta/s)_\mathrm{HG}$ values noted 
earlier (Figs.~\ref{F3}-\ref{F5}) is not caused by the viscous pressure
anisotropy $\Delta$, but by the average transverse viscous pressure $\Sigma$
(green triangles in Fig.~\ref{F8}f). This phenomenon is driven by the 
effect of $\Sigma$ on the proton spectra (Figs.~\ref{F7}c and \ref{F8}c): 
$\Sigma$ suppresses the spectra at low $p_T$, leading (in extreme situations) 
to the formation of a shoulder in the proton spectra which is known 
\cite{Huovinen:2001cy} to cause negative $v_2$.

\section{Large hadronic relaxation times}
\label{sec6}

%
\begin{figure}[b!]
\includegraphics[width=1.0\linewidth,clip=]{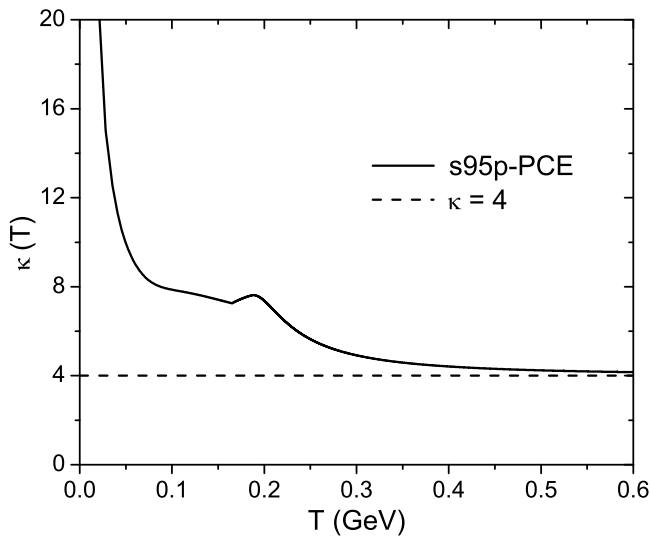}
\caption{\label{F9} The temperature dependent $\kappa(T)$ from 
Eq.~(\ref{eq6}) for EOS s95p-PCE (solid), compared with the massless limit 
$\kappa\eq4$ (dashed).
}
\end{figure}
%

Motivated by the study of the {\tt VISHNU} model in \cite{Song:2010aq}
we explore in this section the consequences of very large relaxation 
times $\tau_\pi$ in the hadronic phase. Specifically, we assume 
a relation proposed in \cite{Koide:2009sy},
\begin{equation}
  \kappa (T) = \frac{e+p}{p}(T),
\label{eq6}
\end{equation}
which can be easily worked out for our EOS s95p-PCE and is shown in 
Fig.~\ref{F9}. In the massless limit (i.e. at large $T$ where the EOS 
approaches $e\eq3p$), this expression approaches the value $\kappa = 4$. 
To explore effects specifically related to the $T$-dependence of $\kappa$, 
we compare in this section results from Eq.~(\ref{eq6}) with those for 
constant $\kappa\eq4$ (and not $\kappa\eq3$ as in the preceding sections).
The QGP viscosity is kept at $(\eta/s)_\mathrm{QGP}\eq0.16$ throughout, 
but we toggle $(\eta/s)_\mathrm{HG}$ in Eq.~(\ref{eq1}) between the two 
values 0.16 and 0.48 (see Fig.~\ref{F1}).

%
\begin{figure}[b]
\includegraphics[width=1.0\linewidth,clip=]{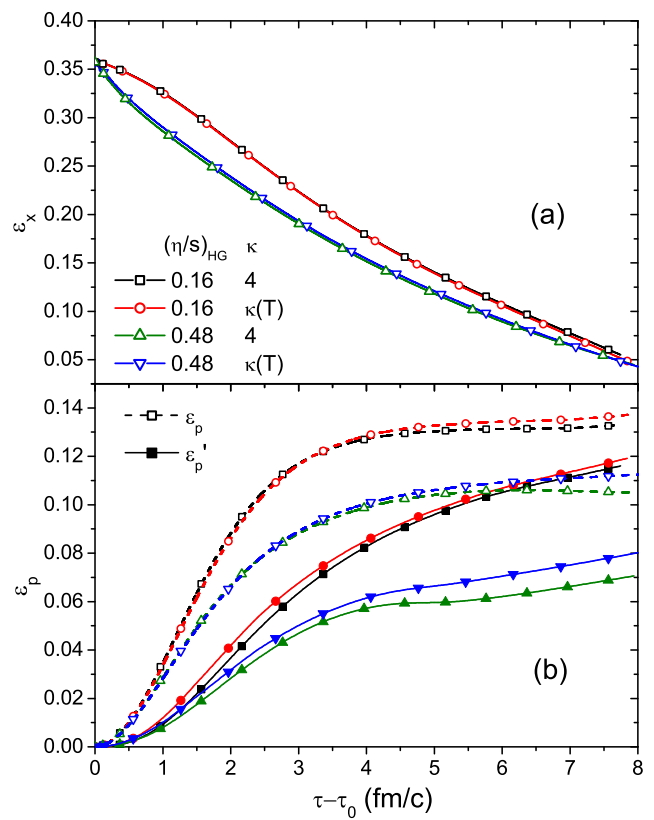}
\caption{\label{F10} (Color online)
Same as Fig.~\ref{F2}b,c, but for $\kappa(T)$ from Eq.~(\ref{eq6}) 
and constant $\kappa\eq4$ instead of $\kappa\eq3$. For the QGP
$(\eta/s)_\mathrm{QGP}\eq0.16$ is used throughout whereas 
$(\eta/s)_\mathrm{HG}$ is varied between 0.16 and 0.48 as indicated 
in the legend.
}
\end{figure}
%

Figure~\ref{F10} shows a similar analysis as Fig.~\ref{F2}, but now
comparing constant with $T$-dependent $\kappa$ values. From Fig.~\ref{F10}a 
we conclude that the temperature dependence of $\kappa$ has no visible
influence on the evolution of the spatial eccentricity $\ecc_x$, irrespective
of whether the specific shear viscosity $\eta/s$ grows in the hadronic phase
or not. On the other hand we see in Fig.~\ref{F10}b that a $\kappa(T)$ that 
grows around and below $\Tc$ as shown in Fig.~\ref{F9} reduces significantly 
the viscous suppression of the total momentum anisotropy $\ecc'_p$ that is 
otherwise caused by a large hadronic shear viscosity.\footnote{Please note 
   that the extremely rapid rise of $\kappa(T)$ below $T\sim 50$\,MeV seen in 
   Fig.~\ref{F9} is irrelevant in this context because the fireball matter 
   decouples already at $\Tdec\eq120$\,MeV.}
Analyzing panel (b) of Fig.~\ref{F10} in more detail, we observe that
during the early stage of the evolution larger hadronic relaxation times 
have little effect on the flow momentum anisotropy $\ecc_p$, consistent 
with the almost unchanged decay rate of the spatial eccentricity seen in 
panel (a) that drives the anisotropic flow. At late times, however, the 
larger $\kappa(T)$ is seen to have a small positive effect on the generation 
of anisotropic collective flow. Increasing the response time $\tau_\pi$ 
with which the viscous pressure tensor $\pi^{\mu\nu}$ can react to
changes in the velocity shear tensor apparantly allows the collective
flow anisotropy to grow more easily, with less viscous damping, than
if $\pi^{\mu\nu}$ is allowed to relax to its Navier-Stokes value 
$\pi^{\mu\nu}_\mathrm{NS}\eq2\eta \sigma^{\mu\nu}$ more quickly.
This is a cumulative effect that becomes visible most clearly at late
times when most of the fireball matter is affected by the larger 
$\kappa(T)$ values at lower temperatures.

%
\begin{figure}[b!]
\includegraphics[width=1.0\linewidth,clip=]{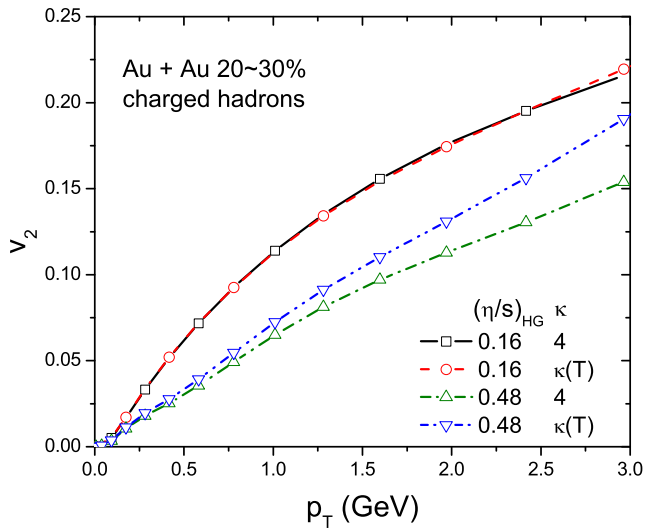}
\caption{\label{F11} (Color online)
Differential elliptic flow $v_2(p_T)$ for charged hadrons, using a
temperature dependent $\kappa(T)$. Same parameters as in Fig.~\ref{F10}.
}
\end{figure}
%

The total momentum anisotropy $\ecc'_p$, on the other hand, is more 
strongly affected by a low-temperature growth of $\kappa(T)$ (solid lines
in Fig~\ref{F10}b). $\ecc'_p$ is suppressed relative to the flow anisotropy 
$\ecc_p$ by the non-equilibrium corrections ${\sim\,}\pi^{\mu\nu}$ in the 
energy-momentum tensor. When the relaxation time $\tau_\pi T$ is allowed 
to grow large in the hadronic phase, this
suppression is found to be reduced, and the reduction is relatively larger
for large values of $(\eta/s)_\mathrm{HG}$ (corresponding to a larger
Navier-Stokes value $\pi^{\mu\nu}_\mathrm{NS}$) than for smaller 
$(\eta/s)_\mathrm{HG}$. We also note that this suppression of the 
$\pi^{\mu\nu}$-contribution to $\ecc'_p$ is visible already at early 
times when the larger $\kappa(T)$ values affect only the fireball 
corona. In fact, for constant $\eta/s\eq0.16$ (solid squares and circles) 
the low-temperature growth of $\kappa(T)$ leads to a bigger increase 
of $\ecc'_p$ over $\ecc_p$ at early than at late times; this is due to the 
larger longitudinal expansion rates at early times which lead to larger 
Navier-Stokes values for $\Delta\eq\pi^{xx}{-}\pi^{yy}$ everywhere, thus 
causing greater sensitivity to increased $\kappa(T)$ values in the fireball 
corona. In the case of $T$-dependent $\eta/s$ (solid upright and inverted 
triangles) the effects from a delayed response $\tau_\pi$ are larger at 
late times; in this situation, the Navier-Stokes values for 
$\Delta\eq\pi^{xx}{-}\pi^{yy}$ grow in the hadronic phase due the sudden 
increase of $\eta/s$ below $T_c$, clearly reflected by a ``kink'' in the 
growth of $\ecc'_p$ around $\tau{-}\tau_0\eq4$\,fm/$c$ (see upright green 
solid triangles in Fig~\ref{F10}b). This kink is largely washed out by a 
simultaneous rise of $\kappa(T)$ (inverted blue solid triangles
in Fig~\ref{F10}b).

The behavior of the total momentum anisotropy $\ecc'_p$ is directly
reflected in the charged hadron elliptic flow, shown in Fig.~\ref{F11}.
We point especially to the reduction of the (negative) $\pi^{\mu\nu}$
contributions to $\ecc'_p$ in the case of $T$-dependent $(\eta/s)(T)$, 
which manifests itself through reduced $\delta f$ corrections to $v_2(p_T)$ 
which again are most pronounced at large $p_T$ (green triangles and blue 
inverted triangles in Fig.~\ref{F11}). For constant $\eta/s$, on the other
hand, the larger hadronic relaxation time has little effect on the 
differential $v_2(p_T)$, consistent with the very small effect 
on the total momentum anisotropy $\ecc'_p$ at late times seen in 
Fig.~\ref{F10}b.

\section{Discussion and conclusions}
\label{sec7}

Figure~\ref{F11} has important implications: Comparing the blue line with 
inverted triangles to the case of constant $\kappa$ and $\eta/s$ (black 
squares), we conclude that the suppression of $v_2(p_T)$ reflected in
the blue line could have arisen in two different ways: (i) by a large
increase of $\eta/s$ in the hadronic phase, accompanied by a similarly 
large increase of $\kappa$, as shown here, or (ii) by a much less 
pronounced increase of the hadronic shear viscosity, compensated by a
correspondingly reduced increase of the hadronic relaxation time. In other
words, the hadronic shear viscosities and relaxation times extracted from a
given charged hadron $v_2(p_T)$ are strongly correlated and impossible to
determine independently from a single elliptic flow measurement. Whether
and how the systematic exploration of differential elliptic flow for
different particle species and different collision systems at different 
centralities can help to resolve this ambiguity remains to be seen.

The study presented here shows that any discussion of large 
dissipative effects in the hadronic phase of heavy-ion collisions,
reflected by specific shear viscosities and (scaled) microscopic
relaxation times that grow as the system cools below the critical
quark-hadron transition temperature, is really a discussion of 
$\delta f$, i.e. of the deviation of the freeze-out distribution function 
from its local equilibrium form and its reflection in the final hadron 
spectra and anisotropies. As the system cools and approaches kinetic
freeze-out, dissipative effects become stronger and stronger, bringing
the framework of viscous hydrodynamics closer and closer to breakdown. 
In this sense, our results have to be taken as qualitative insights but 
should not be confused with quantitative predictions. Their main value, 
as we see it, is that they shed light on and help to classify and 
qualitatively understand the late-stage dissipative effects on hadron 
spectra and their elliptic flow as seen in a realistic microscopic 
approach (as embodied, for example, by \VC). The results presented here
do provide support to the conclusion of Ref.~\cite{Song:2010aq} that
an effective viscous hydrodynamic description of the hadronic stage
in heavy-ion collisions, if valid at all, likely requires both large
shear viscosity {\em and} long relaxation times below $\Tc$.

\acknowledgments
\VISH\ was written by Huichao Song \cite{Song:2009gc}. We thank Pasi 
Huovinen and Huichao Song for clarifying discussions, and Tomoi Koide for 
providing us with tables for the figures in Ref.~\cite{Koide:2009sy} and 
related discussions. We also express our thanks to an anonymous referee
who suggested the analyses presented in Sec.~\ref{sec6} of this work;
a preliminary version, in which we explored a much more crude and 
unrealistic parametrization of the temperature dependence of $\kappa(T)$ 
can be found on the e-print archive \cite{Shen:2011kn}. This work was 
supported by the U.S.\ Department of Energy under contracts 
\rm{DE-SC0004286} and (within the framework of the JET Collaboration) 
\rm{DE-SC0004104}. 
%

\bibliographystyle{h-physrev3}

\bibliography{references}


\end{document}